\documentstyle[preprint,eqsecnum,aps,epsf]{revtex}	

\newif\iftightenlines\tightenlinesfalse
\tightenlines\tightenlinestrue

\begin{document}
%
\def\gsim{\lower.7ex\hbox{$\;\stackrel{\textstyle>}{\sim}\;$}}
\def\lsim{\lower.7ex\hbox{$\;\stackrel{\textstyle<}{\sim}\;$}}
\def\mol{Mol}
\def\etmiss{E\llap/_T}
\def\eslt{E\llap/_T}
\def\esl{E\llap/}
\def\msl{m\llap/}
\def\to{\rightarrow}
\def\te{\tilde e}
\def\tmu{\tilde\mu}
\def\ttau{\tilde\tau}
\def\tl{\tilde\ell}
\def\ttau{\tilde \tau}
\def\tg{\tilde g}
\def\tnu{\tilde\nu}
\def\tell{\tilde\ell}
\def\tq{\tilde q}
\def\tu{\tilde u}
\def\tc{\tilde c}
\def\ts{\tilde s}
\def\tb{\tilde b}
\def\tst{\tilde t}
\def\tt{\tilde t}
\def\tw{\widetilde W}
\def\tz{\widetilde Z}

\hyphenation{mssm}
%
\preprint{\vbox{\baselineskip=14pt%
   \rightline{UM-TH-98-16}\break 
}}
\title{Electric Dipole Moments Do Not Require the CP-violating 
Phases of Supersymmetry To Be Small}
\author{Michal Brhlik, Gerald J. Good and G.L. Kane}
\address{
Randall Laboratory of Physics,
University of Michigan,
Ann Arbor, MI 48109 USA
}
\date{\today}
\maketitle
\begin{abstract}
We report the first fully general numerical calculation of the neutron and 
electron
dipole moments, including the seven significant phases. We find that there are 
major regions in the parameter space where none of the phases are required to 
be small, contrary to the conventional wisdom. The electric dipole moments 
(EDM's) do provide useful constraints, allowing other regions of parameter 
space to be carved away.
We keep all superpartner masses light so agreement with experimental limits 
arises purely from interesting relations among soft breaking parameters.

\end{abstract}

\medskip
\pacs{PACS numbers: ...}



\section{Introduction}

The general parametric structure of the Minimal Supersymmetric Standard Model 
(MSSM) includes a large number of 
CP-violating phases. Their presence has largely been ignored in 
phenomenological analyses because of severe constraints imposed on individual
phases by the experimental upper limits for electron and neutron electric 
dipole moments if other phases are set to zero.
These limits would generally constrain the phases considered individually to be 
less than $10^{-2}$ unless the mass parameters are pushed beyond the 
TeV scale \cite{masi}.
Recently, it has been 
emphasized \cite{nath}, however, that 
cancellations between different terms contributing to the dipole moments
can allow for values of the phases very different from zero even when the 
superpartner masses are relatively light. Since this can have important 
consequences \cite{bk1} for extraction of the parameters in the SUSY Lagrangian 
from experimental data, for calculation of dark matter densities and scattering 
cross sections, for baryogenesis, for Higgs boson limits, and more,
it is rather important to study the problem 
of constraints on the complex phases without making any unnecessary simplifying 
assumptions based on theoretical prejudice. To put it differently, the phases
may or may not actually be small. We must find out from data, without making
assumptions that lead to excluding regions of parameter space where parameters 
are large. {\it If} the phases are large, they affect many CP-conserving
quantities throughout particle physics, so it is even more important to 
proceed carefully. The phases can only be large if certain approximate 
relations among soft breaking parameters hold; these relations will be 
checked in future experiments. The relevant relations are not fine tuned, but
are quite reasonable, with various soft-breaking parameters related in size 
and sign to one another.  

Some important results have already been reported in literature. Nath and 
Ibrahim have presented \cite{nath} some of the formulas needed in the analysis 
and calculated the EDM's in the framework of minimal supergravity model.
Olive {\it et al.} \cite{olive} have analyzed the case of two phases and 
confirmed the analysis of Nath and Ibrahim; they also applied it to the 
calculation of neutralino relic density and detection rate. Some similar 
results were reported in Ref. \cite{bk1} and phenomenological consequences of 
CP-violating phases in the MSSM were studied in \cite{phen}.    

In this paper we want to address this issue in its entirety in order to 
establish a connection between the usual parameters of the MSSM Lagrangian
and ranges of the phases allowed by experimental data on the electric dipoles.
We work in the framework of the simplest possible model neglecting the flavor
mixing but avoiding any assumptions about unification of the soft breaking 
parameters. We use light superpartners masses, so apart from relations among 
soft breaking parameters the resulting EDM's would be very large.

\section{Phase structure of the full MSSM}  
 
We define the MSSM to be the supersymmetric theory with the same particles as 
the Standard Model (SM) plus their superpartners, the SM gauge group, two Higgs 
doublets, and conserved R-parity.

The MSSM Lagrangian \cite{martin,disu} depends on a total of 126 parameters and   
it includes three well known sources of CP-violating phases. The first is 
related to the two Higgs doublets present in the model since both the 
$\mu$ parameter in the superpotential and the soft breaking parameter $b$
can be complex and their phases are denoted $\varphi_{\mu}$ and $\varphi_{b}$
respectively. Three more phases, $\varphi_{1}$, $\varphi_{2}$, and 
$\varphi_{3}$, enter through the complex masses of the gauginos associated with 
the standard gauge groups.
Finally, most of the phases originate in the flavor sector of the Lagrangian, 
either in the scalar soft mass matrices 
$\bf m^2_{Q,\bar{u},\bar{d},L,\bar{e}}$ or the trilinear matrices 
$\bf a_{u,d,e}$. The mass matrices are hermitian so only off-diagonal 
terms can be complex but the trilinear matrices are general $3\times 3$ 
matrices
allowing for the the diagonal entries to also be complex.

The impact of the phases associated with the off-diagonal terms on experimental 
observables is suppressed by the same mechanism which is required to suppress 
the existence of large flavor changing neutral effects and for the purposes of 
this study all these phases can be neglected; if some of them matter it will 
only strengthen our results.
We assume that all the scalar soft
mass matrices and trilinear parameters are flavor diagonal and that the 
complex trilinear terms are proportional to the corresponding Yukawa couplings 
$a_f=y_f A_f e^{i{\varphi_{A_f}}}$, where $\varphi_{A_f}$ are the relevant 
phases \footnote{In further text we take (unless explicitly stated otherwise)
all the dimensionful parameters to be real and positive; their phases are  
always written explicitly.}. 

It is important to realize that not all of the listed phases are physical. 
Above the electroweak symmetry breaking scale, the Lagrangian possesses two 
partial $U(1)$ symmetries which can be promoted to full symmetries by treating
the dimensionful parameters as spurions charged under those symmetries 
\cite{disu}.
Under an R-symmetry the Grassmann variable $\theta$ ($\bar{\theta}$) is charged 
$+1$ ($-1$) and therefore this symmetry distinguishes between 
component fields
of the superfields. If the charge of a chiral superfield is $r$, its scalar 
component field $\phi$ transforms under R-symmetry with charge $r$, the 
fermionic field 
$\psi$ has charge $r-1$ and the auxilary F scalar field possesses charge $r-2$.
In order to preserve the R-invariance of the superpotential it is convenient to 
choose $r=1$ for the matter superfields and $r=0$ for the Higgs superfields.
The advantage of this choice is also clear from the fact that R-symmetry
defined in this way is not broken in the process of electroweak symmetry 
breaking. The vector superfields are not charged under R symmetry, and so only 
the gaugino component field $\lambda$ ($\bar{\lambda}$) obtains a charge of 
$+1$  
($-1$). It is clear that to preserve R symmetry as a full symmetry of the 
superpotential and also of the soft SUSY breaking terms in the Lagrangian,
$\mu$, $A_f$ and the gaugino masses $M_i$ have to be charged under the 
R-symmetry.

The whole MSSM Lagrangian with the exception of the $\mu$-term in the 
superpotential and the $b$ soft breaking term is also invariant under a
Peccei-Quinn  (PQ) symmetry. This symmetry transforms the Higgs fields with 
charge 
$-2$ and the matter fields  $Q,\bar{u},\bar{d},L,\bar{e}$ with charge $+1$.
Again, if $\mu$ and $b$ are treated as spurions full symmetry is restored 
above the electroweak scale. Below this scale, the PQ symmetry is broken as 
the Higgs fields acquire vacuum expectation values. 
Physical observables can only depend on such combinations of parameters which 
are invariant under all symmetries of the Lagrangian. For the unbroken theory 
we have two symmetries and therefore two conditions, allowing us to eliminate 
two phases. When electroweak symmetry is broken we are left with only one 
unbroken symmetry, but the phase of $b$, which is related to the phase of the 
Higgs 
vacuum expectation values, can be absorbed into the physical Higgs fields by
appropriate redefinition. It is therefore natural to take $b$ (and the VEV's)
to be real and set one more phase to be zero. We prefer in this paper to 
take $\varphi_2=0$
thus explicitly violating reparametrization invariance but one has to keep in
mind that all other parameter choices are related to our choice by an 
R-transformation. A fully reparametrization invariant approach to CP-violation 
in SUSY theories will be discussed elsewhere \cite{abks}. Our numerical results 
do not depend in any way on this simplification.
    
Taking into account our parametrization choice, the final set of set of phases
considered in the discussion of the electron and neutron electric dipole 
moments includes three phases appearing in the chargino-neutralino-gluino 
sector, namely $\varphi_1$, $\varphi_3$ and $\varphi_{\mu}$, and four phases 
$\varphi_{A_u}$, $\varphi_{A_d}$, $\varphi_{A_t}$ and $\varphi_{A_e}$ 
corresponding to the trilinear soft breaking parameters relevant in the dipole
moment calculation as discussed in the following section. As we will see below, 
even though $a_u$, $a_d$, $a_e$ are proportional to small Yukawa couplings, 
their phases enter because contributions to the EDM's require a chirality flip 
leading to dipole moments' proportionality to the relevant mass.
  
\section{Electric dipole moment calculation}

The electric dipole interaction of a spin-1/2 particle $f$ with 
an electromagnetic field is described by an effective Lagrangian
\begin{equation}
{\cal L}_{EDM}=-\frac{i}{2} d_f \bar{f} \sigma^{\mu\nu} \gamma_5 f 
F_{\mu\nu}.
\end{equation}   
In theories with CP-violating interactions, the electric dipole $d_f$ receives 
contributions from loop diagrams.  
The best way to account for such contributions is to use the effective theory 
approach in which the heavy particles are decoupled at some large scale $Q$ and
the full theory is matched with an effective theory including a full set of 
CP-violating operators \cite{con1,con2,glu}. If we restrict ourselves to 
dimension 5 and 6 operators the effective Lagrangian takes the form
\begin{equation}
{\cal L}_{eff}=\sum_{i=1}^{3} C_{i}(Q)O_{i}(Q) 
\end{equation}
where the $C_i(Q)$ are Wilson coefficients evaluated at scale $Q$,
and the $O_i$ are the three considered operators
\begin{eqnarray}
O_{1}&=& -\frac{i}{2} \bar{f} \sigma_{\mu\nu} \gamma_5 f 
F_{\mu\nu}, \\
O_{2}&=& -\frac{i}{2} \bar{f} \sigma_{\mu\nu} \gamma_5 T^a f 
G^a_{\mu\nu}, \\ 
O_3&=& -\frac{1}{6} f_{abc} G^a_{\mu\rho} G^{b \rho}_{\ \nu} 
G^c_{\lambda\sigma}
\epsilon^{\mu\nu\lambda\sigma}. \\ 
\end{eqnarray}
It is obvious that all three operators contribute when the 
external fermionic particles are quarks, while in the case of the electron 
$C_2^e$ and $C_3^e$ are identically zero. 

Supersymmetric models contribute to the Wilson coefficients at the one loop 
level and they include several types of graphs as shown in Fig. 1. Chargino, 
neutralino and gluino 
loops where the second particle in the loop is a scalar superpartner, either a 
slepton or a squark, contribute to $C_1$ and $C_2$ coefficients
depending on whether a photon or a gluon is radiated. 
The contributions can be calculated at the electroweak scale since a typical 
SUSY scale in most models is of the same order of magnitude. 
For the gluino loop contribution to the quark EDM the matching gives
\begin{eqnarray}
C_{1}^{q_k-\tilde{g}}(Q)&=& -\frac{2}{3} \frac{e \alpha_S}{\pi}
\sum_{i=1}^6 {\rm Im}(\Delta^{q_k-\tilde{g}}_i) \frac{m_{\tilde{g}}}{m_i^2}
B(\frac{m_{\tilde{g}}^2}{m_i^2}), \\
C_{2}^{q_k-\tilde{g}}(Q)&=& \frac{g_S \alpha_S}{4\pi}
\sum_{i=1}^6 {\rm Im}(\Delta^{q_k-\tilde{g}}_i) \frac{m_{\tilde{g}}}{m_i^2}
C(\frac{m_{\tilde{g}}^2}{m_i^2}). 
\end{eqnarray} 
The neutralino and chargino loops contribute both to the electron and quark  
electric dipole moments and one finds
\begin{eqnarray}
C_{1}^{f_k-\tilde{N}}(Q)&=& \frac{e \alpha}{4\pi\sin^2\theta_W} Q_f
\sum_{i=1}^6 \sum_{j=1}^4 {\rm Im}(\Delta^{f_k-\tilde{N}}_{ij}) 
\frac{m_{{\tilde N}_j}}{m_i^2} B(\frac{m_{{\tilde N}_j}^2}{m_i^2}), \\
C_{2}^{q_k-\tilde{N}}(Q)&=& \frac{g_S g^2}{16\pi^2}
\sum_{i=1}^6 \sum_{j=1}^4 {\rm Im}(\Delta^{q_k-\tilde{N}}_{ij}) 
\frac{m_{{\tilde N}_j}}{m_i^2} B(\frac{m_{{\tilde N}_j}^2}{m_i^2})
\end{eqnarray} 
and
\begin{eqnarray} 
C_{1}^{f_k-\tilde{C}}(Q)&=&-\frac{e \alpha}{4\pi\sin^2\theta_W} 
\sum_{i=1}^6 \sum_{j=1}^2 {\rm Im}(\Delta^{q-\tilde{C}}_{ij}) 
\frac{m_{{\tilde C}_j}}{m_i^2} 
[Q^{\prime}_{f} B(\frac{m_{{\tilde C}_j}^2}{m_i^2})+
 (Q_f- Q^{\prime}_{f}) A(\frac{m_{{\tilde C}_j}^2}{m_i^2})], \\
C_{2}^{q_k-\tilde{C}}(Q)&=& -\frac{g_S g^2}{16\pi^2}
\sum_{i=1}^6 \sum_{j=1}^2 {\rm Im}(\Delta^{q_k-\tilde{C}}_{ij}) 
\frac{m_{{\tilde C}_j}}{m_i^2} B(\frac{m_{{\tilde C}_j}^2}{m_i^2}).
\end{eqnarray}
In equations 3.7-12, $m_i$ are the masses of the corresponding
scalar particle running in the loop and $A$, $B$ and $C$ are the loop functions 
obtained by integrating out the heavy particles in the loop. These functions,
together with the vertex 
$\Delta$ functions calculated in our phase parametrization, can be found in 
the Appendix. $Q_f$ denotes the electric charge of the external fermion and
$Q^{\prime}_{f}$ is the charge of internal sfermion when different from $Q_f$.  

The gluonic operator $O_3$ obtains a contribution from the top-stop loop
with a gluino exchange as shown in Fig. 1 and one has
\begin{eqnarray} 
C_{3}^{f}(Q)&=&-3 \alpha_S m_t (\frac{g_S}{4\pi})^2
{\rm Im}(\Delta^{u_3-\tilde{g}}_2)
\frac{m_{{\tilde t}_1}^2 - m_{{\tilde t}_2}^2}{m_{\tilde g}^5}
H(\frac{m_{{\tilde t}_1}^2}{m_{\tilde g}^2},
  \frac{m_{{\tilde t}_2}^2}{m_{\tilde g}^2},
  \frac{m_t^2}{m_{\tilde g}^2}),
\end{eqnarray}
where the loop function $H$ can be found in the Appendix.

The Wilson coefficients then have to be evolved  from the decoupling scale 
$Q$ down below the chirality breaking scale $\Lambda_{\chi}$ using the 
renormalization group equations (RGE's) in order to account for resummation
of the logarithmic corrections. So far, only QCD corrections for quark 
operator Wilson coefficients  have been estimated \cite{arno} for RGE
evolution down from the electroweak scale to $\Lambda_{\chi}$ giving
\begin{eqnarray}
C_i^q(\Lambda_{\chi})=\eta_i C_i^q(Q),
\end{eqnarray}
where $\eta_1\simeq 1.53$ and $\eta_2\simeq\eta_3\simeq 3.4$. All other 
corrections are neglected in our calculation. 
At the low scale, the CP-violating operators $O_i$ have to be projected on the 
electric dipole operator to evaluate their contribution to the numerical value 
of the electric dipole. This is a complicated task since the chirality 
breaking scale $\Lambda_{\chi}=1.18\, {\rm GeV}$ is very close to the QCD scale
and perturbative methods are not reliable in this region.
The best thing one can do at present is to use naive dimensional analysis 
\cite{mano} which yields
\begin{eqnarray}   
d_f=C_{1}^f (\Lambda_{\chi})+\frac{e}{4\pi} C_{2}^f (\Lambda_{\chi})+
\frac{e \Lambda_{\chi}}{4\pi} C_{3}^f (\Lambda_{\chi}).
\end{eqnarray} 
Finally, since the neutron is a composite particle, one has to use 
a phenomenological neutron model to calculate the neutron EDM
from the moments of the constituting quarks. From the simple $SU(6)$ quark 
model one obtains 
\begin{eqnarray}   
d_n=\frac{1}{3}(4d_d-d_u),
\end{eqnarray}
where $d_d$ and $d_u$ are the EDM's of the down and up quark respectively.

One of the important features of the contributions to the EDM is the fact 
that the effective Lagrangian in Eqn. 3.1 requires different chirality of the
the initial and final particle. In the supersymmetric diagrams this can happen
in two ways --- either the exchanged squark or slepton change chirality via 
L-R mixing terms in the sfermion ${\rm mass}^2$ matrices and couple to the 
gaugino component of the intermediate spin-1/2 particle, or the L and R
sfermions preserve their chirality and couple to the higgsino components of 
charginos or neutralinos. As a result, all contributions are directly 
proportional to the mass of the external particle since both the L-R mixing
sfermion mass term and the higgsino-fermion-sfermion coupling are proportional 
to the relevant Yukawa coupling.   
Another consequence of the chirality flip is the explicit proportionality 
of the contributions to the mass of the intermediate spin-1/2 particle.   

\section{Constraints on the phases}

As already mentioned, we present a numerical treatment
of the electric dipole moment calculation, with the main emphasis on the 
cancellations between various contributions to the Wilson 
coefficients. This allows large values of the phases to give contributions
consistent with the 
experimental bounds on the values of the electric dipole moment of both the 
electron and the neutron. Current experimental limits for the neutron require
that \cite{nexp} 
\begin{eqnarray}   
|d_n|< 1.1 \times 10^{-25} {\ \rm ecm},
\end{eqnarray}   
and for the electron \cite{eexp} 
\begin{eqnarray}   
|d_e|< 4.3 \times 10^{-27} {\ \rm ecm},
\end{eqnarray}
at 95\% confidence level.

We start our analysis by choosing a simple set of the MSSM parameters which 
leads to a fairly light spectrum of the superpartners while still keeping 
the general set of the seven relevant CP-violating phases, $\varphi_1$, 
$\varphi_3$, $\varphi_{\mu}$,  
$\varphi_{A_u}$, $\varphi_{A_d}$, $\varphi_{A_t}$ and $\varphi_{A_e}$, which we 
consider. Since our results do not assume heavy spectrum suppression
of the CP-violation effects, they are fairly general in the sense that 
increasing the masses of the superpartners can only broaden the  
effect of cancellation between different contributions to the electric dipole 
moment. The resulting ranges of the phases for different spectra will differ 
quantitatively from our examples but the general observation that the phases 
indeed do not have to be small will still remain valid. 
To simplify the set of parameters we assume that the squark and slepton soft 
masses and the trilinear soft parameters are flavor diagonal. We also 
take the diagonal entries of these matrices to be universal for all three 
generations and neglect any splitting between up-type and down-type 
right-handed squark masses. Similarly, for the sneutrinos we consider a single
universal soft mass for all three flavors. The trilinear 
parameter $A$ is assumed to be not only flavor universal but also the same for
both sleptons and squarks.
As a result, we are 
left with the following set of soft parameters in the scalar flavor sector
---  $m_{\tilde{\nu}}$, $m_{\tilde{\ell}_L}$, $m_{\tilde{\ell}_R}$
$m_{\tilde{q}_L}$, $m_{\tilde{q}_R}$ and $A$.
We do not assume any relation between the gaugino masses other than taking 
$M_1<M_2<M_3$. Unless stated otherwise, all of our calculations consistently 
employ a common set of parameters shown in the following table:

\begin{center}
\begin{tabular}{|c|c|}
\hline
\multicolumn {2}{|c|} {Standard set of parameters}\\ 
\multicolumn {2}{|c|} {(values at EW scale)}     \\\hline \hline
$M_1=75 {\rm\ GeV}  $ & $m_{\tilde{\nu}}=185 {\rm\ GeV} $ \\ \hline
$M_2=85 {\rm\ GeV}  $ & $m_{\tilde{\ell}_L}=195 {\rm\ GeV} $ \\ \hline
$M_3=250 {\rm\ GeV} $ & $m_{\tilde{\ell}_R}=225 {\rm\ GeV} $ \\ \hline
$\mu=450 {\rm\ GeV} $ & $m_{\tilde{q}_L}=340 {\rm\ GeV} $ \\ \hline
$A=250 {\rm\ GeV}   $ & $m_{\tilde{q}_R}=360 {\rm\ GeV} $ \\ \hline
$\tan\beta=1.2      $ & $m_{A}=300  {\rm\ GeV}            $\\ \hline
\end{tabular}
\end{center}    

We have varied them sufficiently to show that our results are qualitatively 
unchanged for significant regions of these parameters.

\subsection{Electron EDM}

Let us now concentrate on discussing the cancellation mechanism in the two 
cases 
of EDM calculation. The electron EDM limits are more constraining than 
the neutron EDM limits and are also
simpler to study since there are only two large contributions. 
The electron EDM calculation involves only the chargino 
and neutralino contribution to the $C_1$ Wilson coefficient corresponding
to the electric dipole moment operator and only three phases enter the 
calculation, namely $\varphi_{\mu}$, $\varphi_{1}$ and  $\varphi_{A_e}$.
Since the neutralinos are mixtures of both $U(1)$ and $SU(2)$ gauginos and 
both neutral higgsinos, the neutralino contribution includes 
both types of chirality flipping processes --- from the gaugino exchange with 
L-R slepton mixing as well as from the process with gaugino-higgsino mixing and 
requiring no chirality flip in the slepton sector. On the other hand, the 
chargino exchange can only proceed through the latter channel since the 
chargino $SU(2)$ gaugino component only couples to left-handed fields. 
In order for a cancellation between these two contributions to occur, certain 
conditions have to be met. 

The first condition requires that the two 
contributions have opposite sign over at least a subset of the phase parameter 
space. In fact, this requirement is automatically satisfied for contributions 
coming from the gaugino-higgsino mixing diagrams. These contributions, involving 
both charginos and neutralinos, depend on $\varphi_{\mu}$ and have 
opposite sign over the whole range of $\varphi_{\mu}$ due to the fact that the 
$\mu$ parameter enters the neutralino and chargino mass matrices with opposite 
phase. This ``fortunate'' feature can be traced back to the antisymmetry of the 
$SU(2)$ metric $\epsilon$ appearing in the superpotential. 

The neutralino 
contribution of this type could in principle also depend on $\varphi_{1}$,
which would upset the exact anticoincidence of the signs. In practice, the 
dominant part of the contribution comes from the $SU(2)$ gaugino-higgsino 
mixing and the 
effect of $\varphi_{1}$ is constrained to a shift in the phase of the 
neutralino contribution from the gaugino-higgsino interaction.
The gaugino-gaugino diagram for the neutralino contribution, unlike the 
gaugino-higgsino diagrams, involves L-R mixing in the selectron sector. The 
imaginary part of the the relevant phase dependent term is
\begin{eqnarray}   
{\rm Im}(\frac{m^2_{LR}}{m^2_{LL}} N_{1j}^{*2}) \simeq -\frac{1}{m^2_{\tilde{e}}} m_e 
[A_e\sin(\varphi_{1}+\varphi_{A_e})+\mu\tan\beta\sin(\varphi_{\mu}-\varphi_{1})
],
\end{eqnarray}        
where $m^2_{LR}$ and $m^2_{LL}$ are elements of the selectron mass matrix.
In order for this expression to have the oposite sign to the chargino 
contribution, which is negative for $0<\varphi_{\mu}<\pi$ and positive for  
$\pi<\varphi_{\mu}<2\pi$, various possibilities occur depending on the 
relative sizes of $A_e$ and $\mu\tan\beta$. 
For example, in the two limiting cases where $|\mu|\tan\beta>>|A_e|$ and 
$|A_e|>>|\mu|\tan\beta$, 
we get in terms of the phases that modulo $2\pi$ we have to impose 
$\varphi_{\mu}-\varphi_{1} \sim -\varphi_{\mu}$ and 
$\varphi_{1}+\varphi_{A_e} \sim -\varphi_{\mu}$ respectively. In general this
contribution is opposite in sign to the chargino one over a significant part 
of parameter space.

The second condition of cancellation requires the chargino and neutralino 
contributions to be of the same magnitude. In the chargino sector, the 
gaugino-higgsino mixing involves only one type of gaugino while in the 
neutralino sector there are two gaugino states and therefore the elements
of the neutralino diagonalizing matrix $N$ generally yield smaller imaginary 
parts than the elements of the chargino matrices $U,V$.   
Moreover, the chargino contribution is enhanced due to larger values of the 
$A(x)$ loop function
as compared to the $B(x)$ function in the neutralino expression. This comes 
from the fact that the photon in the chargino loop diagram is emitted from the 
fermionic leg of the corresponding diagram as opposed to the scalar leg in the 
neutralino diagram. Both these effects decrease the relative magnitude of the 
neutralino contribution compared to the chargino contribution.
On the other hand, in the neutralino case the gaugino-gaugino contribution can 
balance some of the difference between the two contributions. For that to 
happen, it is important that the relative size of the chargino-higgsino 
contribution decreases and the relative size of the gaugino-gaugino 
contribution increases, which can be achieved by increasing $\mu$. That brings 
us back to the first condition of opposite sign which can be satisfied for 
$|\mu|\tan\beta>>|A_e|$ if $\varphi_1\sim\pi$.  

As a result, for suitable combinations of the dimensionful parameters an 
almost exact cancellation can occur for the whole range of $\varphi_{\mu}$ as 
exemplified in Fig.2{\it a}. In this plot we chose $\mu=700 {\rm\ GeV}$ in addition 
to our standard set of parameters and $\varphi_1$ was set to be equal to 
$\pi$. The values of $\varphi_{A_e}$ were varied randomly leading to the result 
that the values of the neutralino contribution and of the total dipole moment 
form  bands of non-zero width, while the chargino contribution is independent 
of $\varphi_{A_e}$. It is clear from the plot, however, that virtually 
all values of $\varphi_{\mu}$ would be allowed for this particular set of 
parameters depending only on a suitable choice of the $\varphi_{A_e}$ value 
range. This is also significant because $\varphi_{A_e}$ is 
otherwise irrelevant not only in the neutron EDM calculation but also in most 
other phenomenological considerations. Later figures show effects of varying
$\mu$. Note that without cancellations one would have to have each contribution
reduced by $\sim 10^{-2}$, i.e. each phase would have to be $\lsim 10^{-2}$,
as in the usual result.     
   
Thus we see that the strong constraint from the electron EDM limit is naturally 
satisfied over a significant part of the parameter space, though not all of it.
The neutralino and chargino contributions can automatically have opposite sign 
and the same magnitude for most of the $\varphi_{\mu}$ range when the mass 
parameters are in certain ratios depending on the other phases. While the 
cancellations do require related magnitudes of some parameters, and could thus 
be interpreted as a fine-tuning, we think that the required mass relations are 
typically the kind of relations that might arise in a theory of the 
soft-breaking parameters, and are likely to be a clue to the form of the 
theory. The resulting relations are 
predictions that can be tested in other experiments.

To put it in another way, there are two ways to satisfy the electron EDM 
constraints. One possibilty is that the phases are small or zero as a result of 
some presently unknown mechanism. 
Alternatively, the phases could be large, and the masses could 
have certain approximate ranges of reasonable values. The relevant signs would 
automatically give the needed cancellation, which need not have happened.
The two alternatives lead  to very different predictions  for many other 
observables. The naturalness of the cancellation that occurs leads us to 
consider the solution with large phases seriously enough to convince us to 
analyze the full parameter space and to study the resulting predictions, which 
we will report on later.

\subsection{Neutron EDM}

Next we turn to the neutron EDM, where cancellations are easier to obtain.
First of all, all three operators in Eqn. 3.2 receive contributions from the
MSSM one loop diagrams involving quarks as incoming and outgoing particles.
The gluino-squark diagram projects on both $O_1$ and $O_2$ operators, and the 
contribution of the relevant Wilson coefficients $C_{1}^{q_k-\tilde{g}}$ and 
$C_{2}^{q_k-\tilde{g}}$ to the EDM is numerically comparable.
The contribution of $C_{2}^{q_k-\tilde{g}}$ is seemingly suppressed by the 
factor of $\frac{e}{4\pi}$ in Eqn. 3.15 compared to $C_{1}^{q_k-\tilde{g}}$, 
but that is compensated by enhancements from the factor of $\frac{g_S}{e}$, 
and mainly from the loop function $C(x)$ in the matching conditions 3.7 and 
3.8. This is again a consequence of the fact that the gluino leg in the
diagram can emit gluons but not photons. 
The chargino loop $C_{1}^{q_k-\tilde{C}}$ contribution is typically of the 
same order as the gluino loop contributions while $C_{2}^{q_k-\tilde{C}}$ 
contributes negligibly since in this case the $\frac{g_S}{e}$ enhancement alone
does not overcome the suppression from $\frac{e}{4\pi}$.
Both neutralino contributions from $C_{1}^{q_k-\tilde{N}}$ and 
$C_{2}^{q_k-\tilde{N}}$ can be safely neglected in the neutron dipole analysis.
Reasons similar to those for the electron case lead to the suppression of 
$C_{1}^{q_k-\tilde{N}}$ compared to $C_{1}^{q_k-\tilde{C}}$, but in the quark 
case this effect is more pronounced since the squarks are typically heavier 
than the sleptons and they have fractional charges. Correspondingly, the 
contribution from $C_{2}^{q_k-\tilde{N}}$ is even smaller than that from 
$C_{1}^{q_k-\tilde{N}}$. This effectively reduces the number of phases by 
eliminating  $\varphi_{1}$ as one of the parameters numerically relevant
in the neutron EDM calculation.

As in the electron case, it is necessary that the chargino 
contribution be opposite in sign to the sum of the other three contributions
for the cancellation to occur. The gluino contribution exhibits the same 
behavior as the gaugino part of the neutralino contribution in the electron 
case and Eqn. 4.3  transforms into
\begin{eqnarray}   
{\rm Im}(\frac{m^2_{LR}}{m^2_{LL}} G^{*2}) \simeq -\frac{1}{m^2_{\tilde{e}}} m_q 
[A_q\sin(\varphi_{3}+\varphi_{A_q})+\mu f(\beta)\sin(\varphi_{\mu}-\varphi_{3})
],
\end{eqnarray}  
where $A_q=A_u, A_d$ and $f(\beta)=\cot\beta,\tan\beta$ for up and down type 
quarks respectively. The contribution from the pure gluonic operator ,
on the other hand, depends only on 
$\varphi_{3}$, $\varphi_{\mu}$ and $\varphi_{A_t}$ as far as phases are 
concerned and the role of $\varphi_{\mu}$ and $\varphi_{A_t}$ is again
determined by the relative size of $\mu\cot\beta$ and $A_t$.
This implies that  
$\varphi_{3}$ and $\varphi_{\mu}$ are the crucial phases in the EDM calculation.
In order to demonstrate the cancellation on a practical example, in Fig. 
2{\it b}
we set $\varphi_3$ equal to $\pi$ and take all three trilinear parameter phases
to be consistent with zero. All these choicess are enforced within a small
variation around the central value leading to a non-zero width of the gluino 
and pure gluonic contribution. In addition we choose $\mu=300\, {\rm GeV}$ 
so that it is comparable in magnitude to $A_q=250\, {\rm GeV}$ 
and the off diagonal squark mixing terms get a comparable contribution from 
both terms in Eqn. 4.4. The resulting sum total of the neutron EDM is consistent 
with zero over a wide range of $\varphi_{\mu}$. As $\varphi_{3}$, 
$\varphi_{A_u}$, $\varphi_{A_d}$ and $\varphi_{A_t}$ are varied this situation
will persist for large but correlated ranges of these phases. The variation 
with $\mu$ is shown in later figures.

\subsection{Numerical results}

The effects of the cancellation mechanism on the ranges of phases allowed by
the EDM experimental limits can be explored by varying all phases randomly
for a given set of mass parameters and plotting the allowed points projected 
on planes in the phase parameter space. In Fig. 3{\it a} and 3{\it b} we show 
the allowed
regions for the standard parameter set with $\mu=450\, {\rm GeV}$ in the 
$\varphi_{\mu}-\varphi_{1}$ and $\varphi_{\mu}-\varphi_{3}$ plane respectively.
The filled black circles signify the points allowed by electron EDM constraints
and the open circles stand for those allowed by the neutron EDM limits.
The  $\varphi_{1}$ dependence has little significance as far as the neutron 
constraints are concerned, while in the electron case $\varphi_{1}$ has to be 
correlated with $\varphi_{\mu}$ in order to satisfy the limits. Only a selected
band of the values of $\varphi_{\mu}$ is allowed by the electron constraints
and the neutron constraint imposes a correlation between the 
values of $\varphi_{3}$ and $\varphi_{\mu}$ within this band.
Still, when these conditions are satisfied,  values of $\varphi_{\mu}$ very 
different from $0$ or $\pi$ are allowed. All values of $\varphi_{1}$
and $\varphi_{3}$ can occur while the EDM limits are respected.

In Fig. 4 we display the same results as in Fig. 3 but we take 
$\mu=60\, {\rm GeV}$. The range of $\varphi_{\mu}$ is constrained by both 
electron and neutron limits in this case, and the interval allowed by both is 
significantly narrower than in the previous case. Nevertheless, all values
of $\varphi_{1}$ and $\varphi_{3}$ are permitted again.  
 
It is important to see how the range of allowed values of $\varphi_{\mu}$ 
depends on $\mu$ because $\varphi_{\mu}$ plays a crucial role in the electron 
as well as in the neutron EDM calculation. Fig. 5 a displays this range for 
both calculations with the standard parameter set and varying $\mu$. The 
overall trend shows that for larger values of  $\mu$ it is easier to satisfy 
the EDM limits. The {\it b} frame shows the effects of $A_e$ variation on the 
electron EDM constraints when $\mu=450\, {\rm GeV}$. Similarly, in Fig. 6 we 
examined the
dependence of the allowed $\varphi_{\mu}$ range on an overall scaling parameter
$x$ which rescales all the dimensionful parameters in the standard set and
$\mu=450\, {\rm GeV}$ according  to the formula $M^{\prime}=x M$. It is 
interesting to note that in order to allow the full range of $\varphi_{\mu}$
one has to go to fairly large parameters $x>4$ while the same 
effect can be obtained by raising $\mu$ to be larger than $450\, {\rm GeV}$.

Finally, in Fig. 7{\it a} and {\it b} we plot the lightest neutralino mass 
vs. $\varphi_{\mu}$ and $\varphi_{1}$ respectively for the standard parameter 
set and $\mu$ varied from $50\, {\rm GeV}$ to $800\, {\rm GeV}$. 
The neutralino masses can vary quite 
dramatically in the allowed regions and this fact substantially affects 
phenomenological observables at colliders and cosmological implications of the 
supersymmetric model. 
 
\section{Conclusion} 
  
We have shown that the role of the cancellation mechanism in the calculation
of the electron and neutron electric dipole moments within the general 
framework of the MSSM including a non-restricted set of CP-violating phases
has crucial consequences for the range of individual phases.
Even with a light sparticle spectrum, phases can have values very different 
from zero and still satisfy experimental bounds on the values of the electron 
and neutron EDM's.

A trivial but possible way to avoid constraints from the dipole moment 
measurements is the traditional one that all supersymmetric phases are equal 
to zero or unnaturally small. This would 
require the existence of some presently unknown mechanism which would ensure 
that there is negligible CP violation in the SUSY breaking sector of 
the MSSM Lagrangian. On the other hand, we have found that the 
phases may be large while certain approximate relations hold among 
the mass parameters and phases, resulting in cancellations in the calculation 
of the electron 
and neutron EDM. These relations could in principle also come from a theory
of SUSY breaking predicting the exact form of the soft SUSY breaking sector
in the Lagrangian. We have shown in this paper that the latter possibility is 
legitimate and the 
ultimate decision between the two alternatives should be made based on 
experimental measurements.

We have presented a study of the constraints imposed on the phases by electron 
and neutron EDM data for some particular values of soft parameters 
with relatively light spectra.
The results exhibit general features typical for similar choices
and they show that all considered phases can have non-zero values. 
$\varphi_{\mu}$ is  
severely constrained while other
phases can have any value as long as certain correlations with $\varphi_{\mu}$
are respected. The constraints on the phases relax as heavier spectra or large
values of $\mu$ are considered.

The fact that phases can be non-vanishing is very important if one considers 
the general correspondence between the parameters in the supersymmetric 
Lagrangian and various observables which will possibly be measured at future 
collider experiments. For example, without a determination of the phases
it is not possible to measure the value of $\tan\beta$. It is also important
to realize that the presence of 
phases has a substantial impact on the neutralino relic density calculation 
and on the magnitude of the corresponding neutralino scattering cross section    
for dark matter detection. If progress in supersymmetric particle physics 
proceeds by the historical path, it will be essential to measure the phases
to learn the form of the soft-breaking Lagrangian, and thereby be led to 
recognize the mechanism of supersymmetry breaking.


\acknowledgments
We are grateful to Toby Falk, Robert Garisto, Howie Haber, Steve Martin and 
Xerxes Tata for valuable discussions. We also thank Dan Chung and Lisa Everett
for helpful conversations and comments on the manuscript.  

\section*{Appendix}

\def\theequation{A.\arabic{equation}}
\setcounter{equation}{0}

In this appendix we summarize all the calculational details necessary for 
evaluation of the contributions to electric dipole moments of elementary 
particles in the MSSM at the one-loop level. The effect of CP-violating phases 
enters through the particular vertex $\Delta$ functions characteristic for each
type of contributing diagrams. These functions depend on the matrices 
diagonalizing the generally complex mass matrices of the participating 
supersymmetric particles.

In our parametrization, the gluino mass is complex and can be diagonalized by 
a single complex number $G$ defined by     
\begin{eqnarray}   
G^* M_3 e^{i \varphi_3} G^{-1}=M_3,
\end{eqnarray}
resulting in $G=e^{i \varphi_3/2}$. Simlarly, the chargino mass matrix
\begin{eqnarray}
{\cal M}_C=\left(\begin{array}{cc}
 M_2 &  \sqrt{2} M_W \sin\beta   \\
\sqrt{2} M_W \cos\beta   & \mu e^{i\varphi_{\mu}}  \end{array}\right)
\end{eqnarray}
is diagonalized by two generally complex unitary matrices $U$ and  $V$
so that 
\begin{eqnarray}
U^* {\cal M}_C V^{-1}={\cal M}_C^{diag}.
\end{eqnarray}
The neutralino mass matrix
contains two phases, $\varphi_{1}$ and $\varphi_{\mu}$, in our parametrization  
\begin{eqnarray}
{\cal M}_N=\left(\begin{array}{cccc}
 M_1 e^{i\varphi_{1}} & 0 & -M_Z \sin\theta_W\cos\beta & M_Z \sin\theta_W
 \sin\beta  \\
 0 & M_2 & M_Z \cos\theta_W\cos\beta & -M_Z \cos\theta_W\sin\beta  \\
 -M_Z \sin\theta_W\cos\beta & M_Z \cos\theta_W\cos\beta & 0 & -\mu 
 e^{i\varphi_{\mu}} \\
 M_Z \sin\theta_W \sin\beta & -M_Z \cos\theta_W\sin\beta & -\mu 
 e^{i\varphi_{\mu}} & 0 \end{array}\right)
\end{eqnarray}
and the diagonalization matrix $N$ satisfies
\begin{eqnarray}
N^{-1*} {\cal M}_N N={\cal M}_N^{diag}.
\end{eqnarray}
Finally, the scalar superpartners of the three families of fermions in the 
Standard model obtain masses through a general mass matrix 
\begin{eqnarray}
{\cal M}_{\tilde{u},\tilde{d},\tilde{e}}=\left(\begin{array}{cc}
 {\bf m^2_{Q,Q,L}} + {\bf m_{u,d,e} m_{u,d,e}^{\dag}} +D_L\bf {1}  & 
 {\bf a_{u,d,e}^{\dag}} v_{u,d,d}-\mu e^{i\varphi_{\mu}} v_{d,u,u}{\bf 1} \\
 {\bf a_{u,d,e}} v_{u,d,d}-\mu e^{-i\varphi_{\mu}} v_{d,u,u}{\bf 1}   & 
 {\bf m^2_{\bar{u},\bar{d},\bar{e},}} + {\bf  m_{u,d,e}^{\dag} m_{u,d,e}}
 +D_R\bf {1} 
 \end{array}\right)
\end{eqnarray} 
where $D_L=M_Z^2 (T_3-Q \sin\theta_W^2) \cos 2\beta $ and $D_R=M_Z^2 
Q \sin\theta_W^2 \cos 2\beta $, and $v_u$, $v_d$ are the VEV's of the two 
neutral Higgs fields coupling to the up-type and down-type particles 
respectively. The parameters in bold print are $3\times 3$
matrices, generally complex as discussed in the main text.
The mass matrix  in Eqn. A.6 can be diagonalized by a pair of $3\times 6$
matrices relating the interaction and mass eigenstates
\begin{eqnarray}
{\tilde{f}}^L_i=\Gamma^L_{(f)ij} {\tilde{f}}^{diag}_j \\
{\tilde{f}}^R_i=\Gamma^R_{(f)ij} {\tilde{f}}^{diag}_j
\end{eqnarray} 
for each type of fermion and all families $i=1,2,3$. Our notation distinguishes 
between the three types of sfermions, $\tilde{u}$ $\tilde{d}$ and $\tilde{e}$,
and individual flavor states are numbered according to the family number, so, 
for example, the ${\tilde{u}}^L_3$ field corresponds to the left-handed top 
squark field.
   
The gluino vertex function reflects the fact that the gluino is a pure gaugino 
and the only possible way to produce a chirality changing effective vertex is 
to make use of L-R squark mixing and get
\begin{eqnarray}
\Delta^{q_k-\tilde{g}}_i= \Gamma^R_{(q)ki} \Gamma^{L*}_{(q)ki} {G^*}^2
\end{eqnarray}   
with no summation implied over $i$ and $q=u,d$. The neutralino vertex function
can be obtained in a similar way giving
\begin{eqnarray}
\Delta^{f_k-\tilde{N_j}}_i&=&\{\sqrt{2} \tan\theta_W Q N_{1j}^* 
\Gamma^R_{(q)ki}-\lambda_f N_{hj}^* \Gamma^{L}_{(q)ki}\}\times \cr   
& &\{-\sqrt{2} [\tan\theta_W (Q-T_3) N_{1j}^*+T_3 N_{2j}^*] \Gamma^{L*}_{(q)ki}
+  N_{hj}^* \Gamma^{R*}_{(q)ki}\}
\end{eqnarray}
where $\lambda_u=\frac{m_u}{\sqrt{2}M_W\sin\beta}$,
$\lambda_d,e=\frac{m_{d,e}}{\sqrt{2}M_W\cos\beta}$, and $h=3$ for $h=d,e$ and 
$h=4$ for $f=u$. It is obvious from the structure of the function that the 
neutralino effective vertex includes both gaugino and higgsino interactions.
Finally, the chargino vertex function for individual types of particles takes 
the form 
\begin{eqnarray}
\Delta^{u_k-\tilde{C_j}}_i&=& \lambda_u V_{j2}^*\Gamma^{L}_{(d)ki} 
(U_{j1}^*\Gamma^{L*}_{(d)ki}-\lambda_d U_{j2}^*\Gamma^{R*}_{(d)ki})\\
\Delta^{d_k-\tilde{C_j}}_i&=& \lambda_d U_{j2}^*\Gamma^{L}_{(u)ki} 
(V_{j1}^*\Gamma^{L*}_{(u)ki}-\lambda_d V_{j2}^*\Gamma^{R*}_{(u)ki})\\
\Delta^{e_k-\tilde{C_j}}_i&=& \lambda_e U_{j2}^*V_{j1}^*.
\end{eqnarray}

In order to make this paper self contained, we also list the necessary loop 
functions coming from integrating out the supersymmetric particles in the one 
loop diagrams in the case of the electric and chromoelectric dipole operators
\cite{fcta}
\begin{eqnarray}
A(x)&=&\frac{1}{2 (1-x)^2} (3-x+\frac{2ln(x)}{1-x})\\
B(x)&=&\frac{1}{2 (1-x)^2} (1+x+\frac{2xln(x)}{1-x})\\
C(x)&=&\frac{1}{6 (1-x)^2} (10x-26+\frac{2 x ln(x)}{1-x}-\frac{18 ln(x)}{1-x})
\end{eqnarray}
and from the two loop calculation in the case of the purely gluonic operator
\cite{fctb}
\begin{eqnarray}
H(z_1,z_2,z_3)=\frac{1}{2} \int_0^1 dx\int_0^1 du \int_0^1 dy\, x (1-x) u 
\frac{N_1 N_2}{D^4},
\end{eqnarray}
where
\begin{eqnarray*}
N_1&=&u(1-x)+z_3 x (1-x) (1-u)-2 ux [z_1 y+z_2(1-y)]\\
N_2&=&(1-x)^2(1-u)^2+u^2-\frac{1}{9} x^2(1-u)^2  \\
D&=&u(1-x)+z_3 x (1-x) (1-u)+ux[z_1 y+z_2 (1-y)].
\end{eqnarray*}
The integrals in the above definition of $H$ can be simplified and evaluated 
numerically.
%

\newpage

%
%
\begin{figure}
\caption[]{
One loop Feynman diagrams contributing to the calculation of the electric 
dipole moments in the MSSM. The gluon and photon line can originate on any
internal leg carrying corresponding charge.}
\label{fig1}
\end{figure}
\begin{figure}
\caption[]{
Illustration of the cancellation mechanism in the EDM calculation.
See the discussion in the text. Frame {\it a}
includes the contributions to the electron dipole moment arising from 
neutralino and chargino loops contributions to the $C_1$ Wilson coefficient 
for varying $\varphi_{\mu}$, $\varphi_{1}\sim \pi$ and values of 
$\varphi_{A_e}$ sampled randomly. 
A standard set of parameters (see text) with $\mu=700\, {\rm GeV}$ was used. 
Frame {\it b} shows the neutron EDM contribution from the gluino loop graph  
projection into $C_1$ and $C_2$ ($\tilde{g_1}$ and $\tilde{g_1}$, from the 
chargino loop contribution to $C_1$ ($\tilde{C_1}$)
and from the gluino-top-stop graph contributing through the purely gluonic
operator Wilson coefficient $C_3$ ($G$). In this case, the standard set of 
parameters is adopted with $\mu=300\, {\rm GeV}$, $\varphi_{3}\sim \pi$ and
$\varphi_{A_q}\sim 0$ for $q=u,d,t$. In both cases the natural cancellations 
can give  a total of order the experimental limits for most or all of 
$\varphi_{\mu}$. If the cancellation effects were not included one would 
conclude that all phases would have to be of order $10^{-2}$ to not exceed 
the experimental limits.  
}   
\label{fig2}
\end{figure}
\begin{figure}
\caption[]{
Plots of regions allowed by the electron (filled circles) and neutron (open
circles) EDM limits in the $\varphi_{\mu}-\varphi_{1}$ plane (frame {\it a})
and the $\varphi_{\mu}-\varphi_{3}$ plane (frame {\it b}). A value
of $\mu=450\, {\rm GeV}$ was chosen together with the standard parameter set 
and all phases were sampled randomly.
}
\label{fig3}
\end{figure}
\begin{figure}
\caption[]{
Same as Fig. 3, but for $\mu=60\, {\rm GeV}$ and the standard set of 
parameters. Again, all phases were varied randomly.
}
\label{fig4}
\end{figure}
\begin{figure}
\caption[]{
Frame {\it a} shows variation of the $\varphi_{\mu}$ allowed region with $\mu$
for the standard set of parameters and other phases sampled randomly.
The values of $A=A_e=A_u=A_d=A_t$ were also varied from $-500\,{\rm GeV}$ to
$500\,{\rm GeV}$. Open (full) circles denote points allowed by the neutron 
(electron) EDM limit.
Frame {\it b} demonstrates variation of the $\varphi_{\mu}$ range allowed by 
the  electron EDM limits with the values of $A$ for $\mu=450\, {\rm GeV}$.
}
\label{fig5}
\end{figure}
\begin{figure}
\caption[]{
We plot the points allowed by the electron EDM limits for parameter sets 
with all the mass parameters scaled by $x$ with respect to the standard set.
All phases are sampled randomly.
}
\label{fig6}
\end{figure}
\begin{figure}
\caption[]{
Plots of the lightest neutralino masses allowed by the neutron (open circles) 
and electron (filled circles) EDM limits vs. $\varphi_{\mu}$ in frame 
{\it a} and $\varphi_{1}$ in frame {\it b}. In addition to all phases, the 
values of $\mu$ were also varied from  $50\,{\rm GeV}$ to
$800\,{\rm GeV}$ and all other parameters were standard. 
}

\label{fig7}
\end{figure}
%

\centerline{\epsfbox{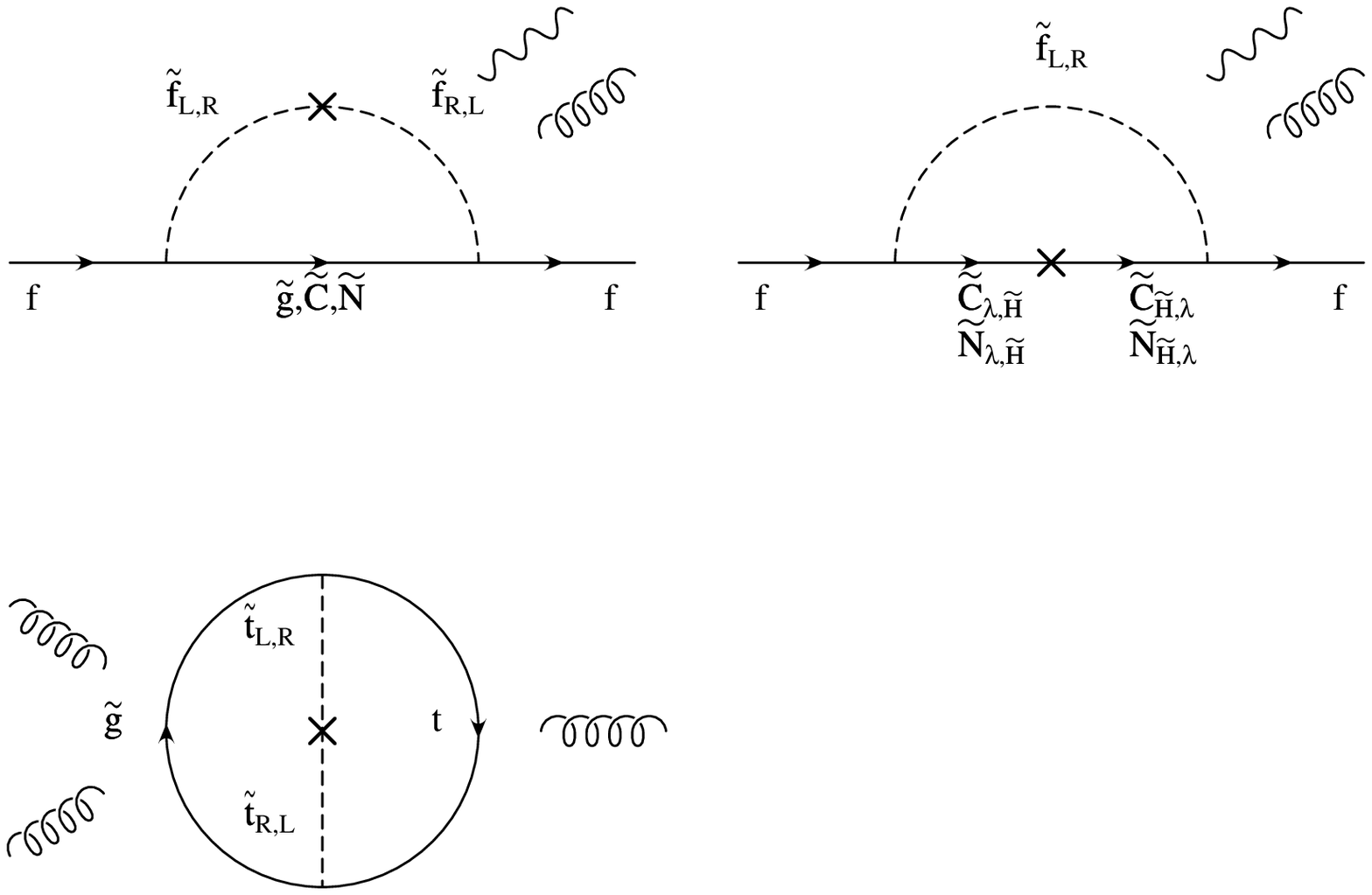}}
\smallskip
\centerline{Fig.~1}
\vfill\eject

\centerline{\epsfbox{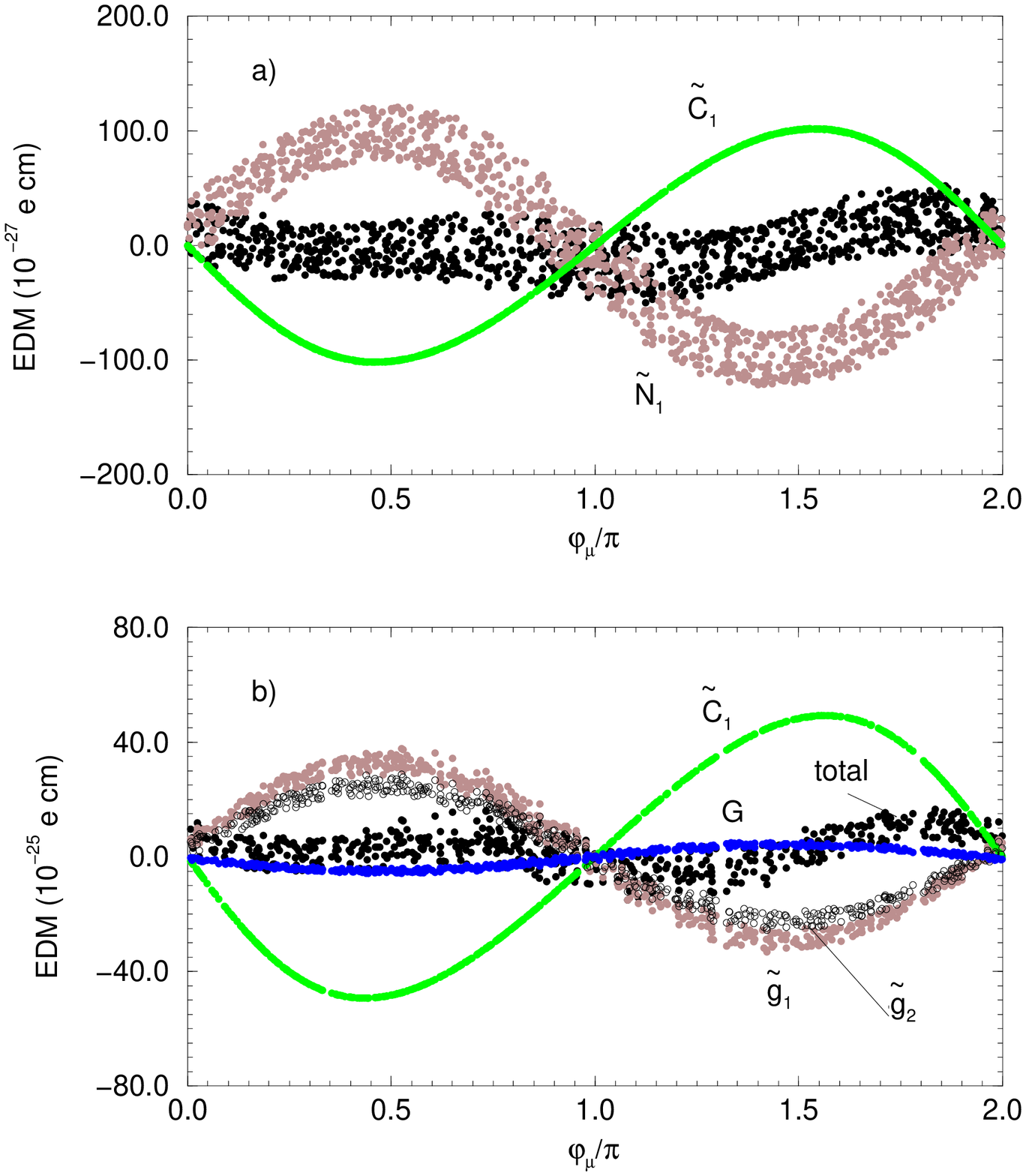}}
\vskip -5mm
\centerline{Fig.~2}
\vfill\eject

\centerline{\epsfbox{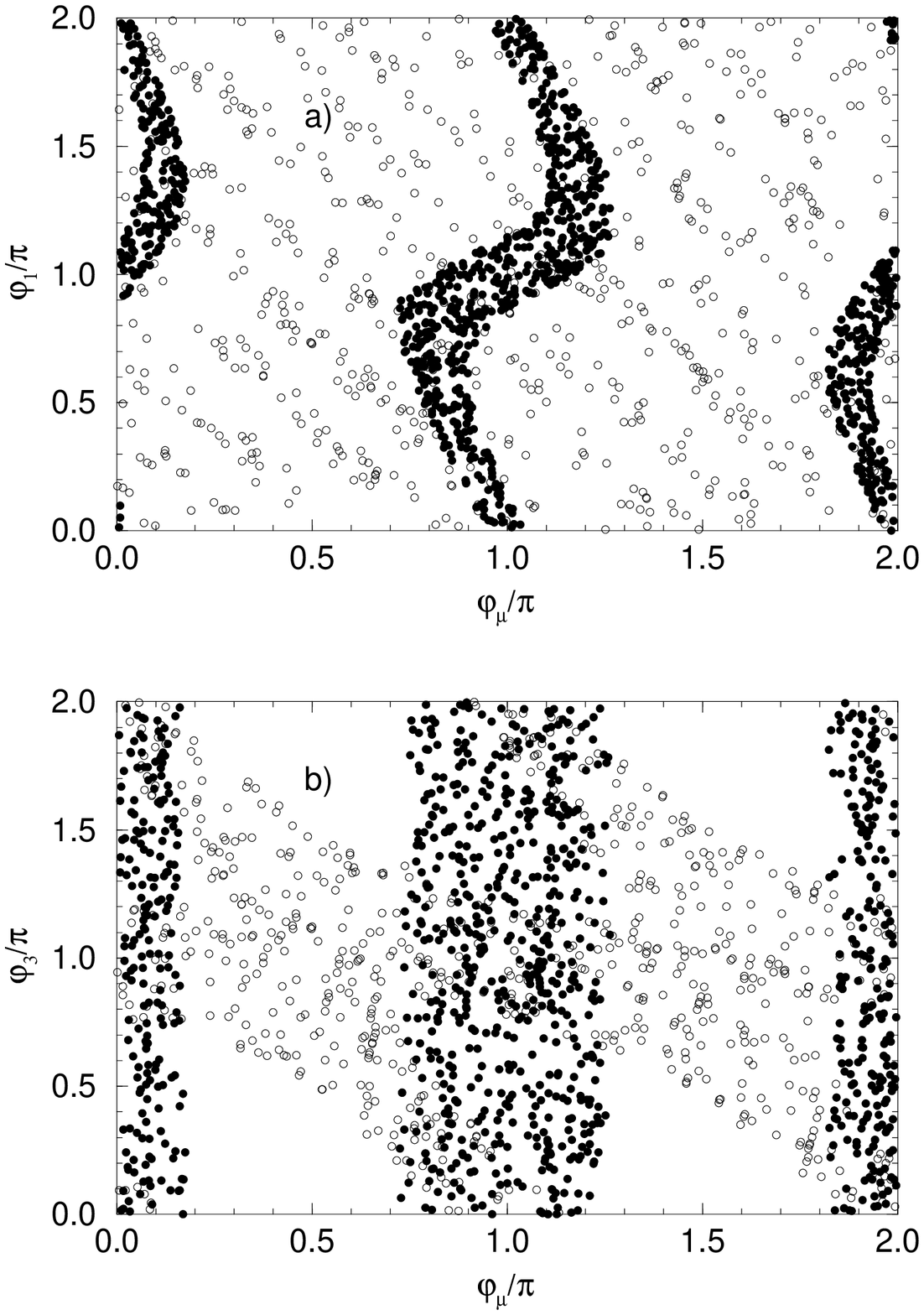}}
\vskip -5mm
\centerline{Fig.~3}
\vfill\eject

\centerline{\epsfbox{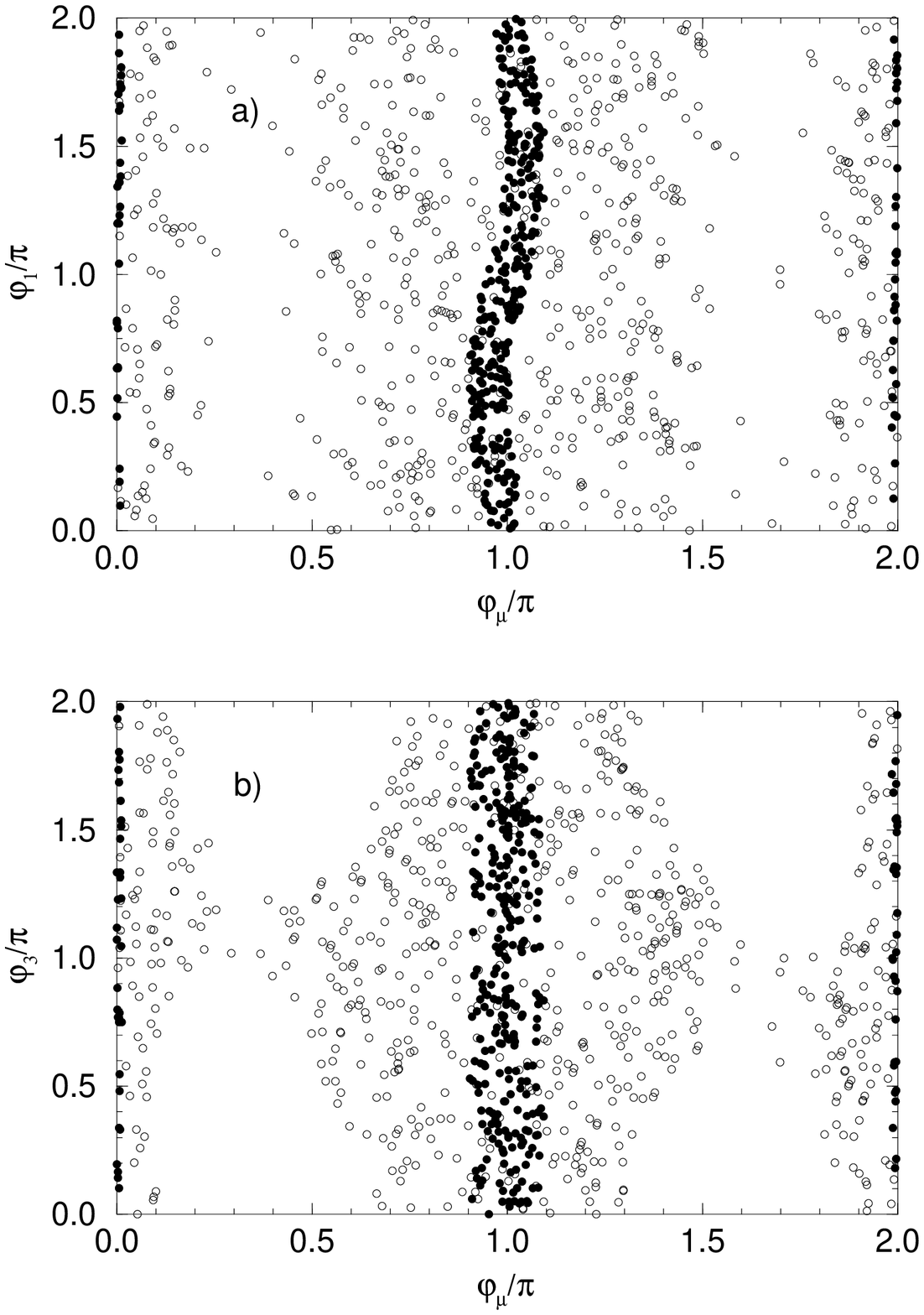}}
\vskip -5mm
\centerline{Fig.~4}
\vfill\eject

\centerline{\epsfbox{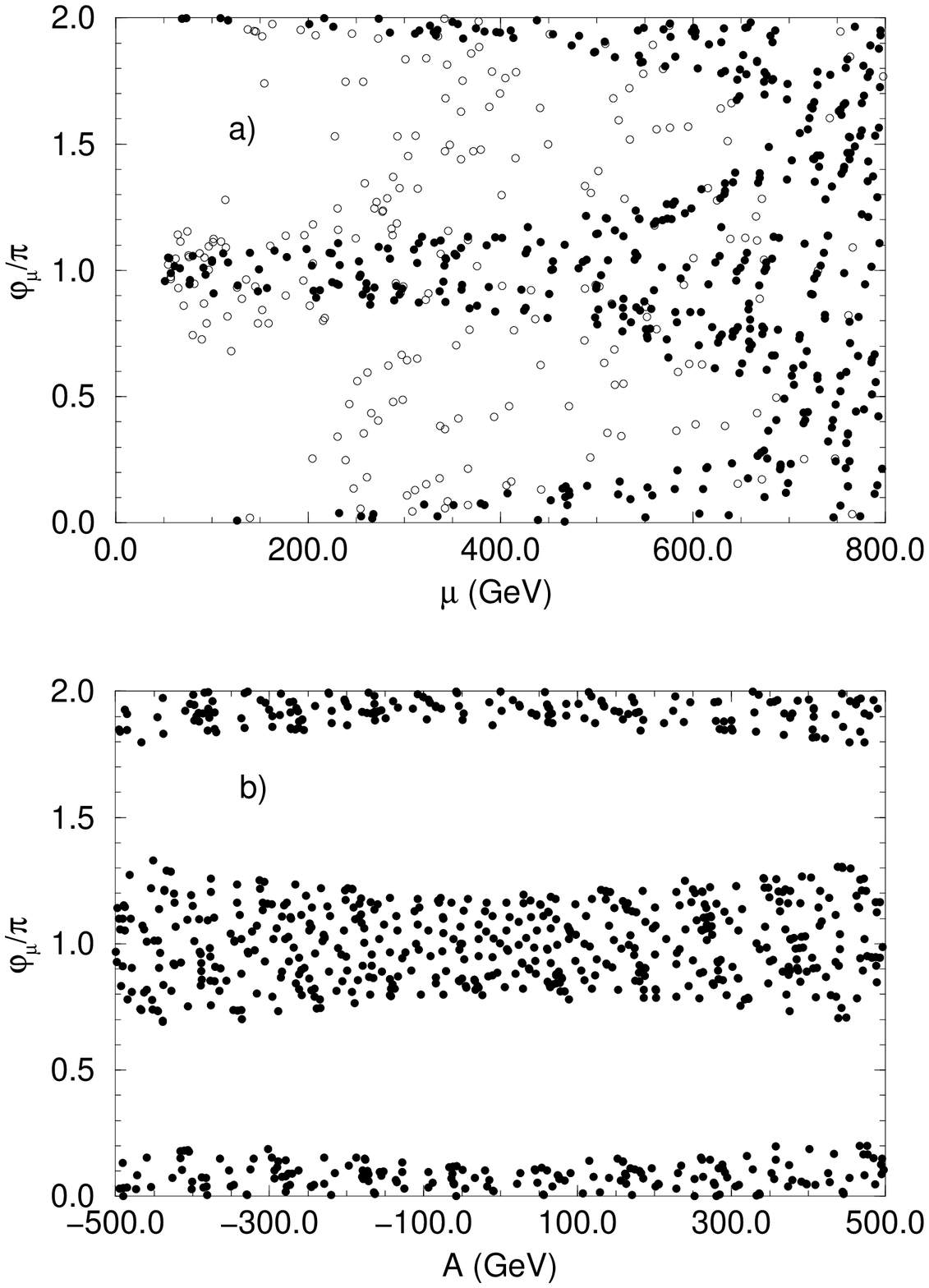}}
\vskip -5mm
\centerline{Fig.~5}
\vfill\eject

\centerline{\epsfbox{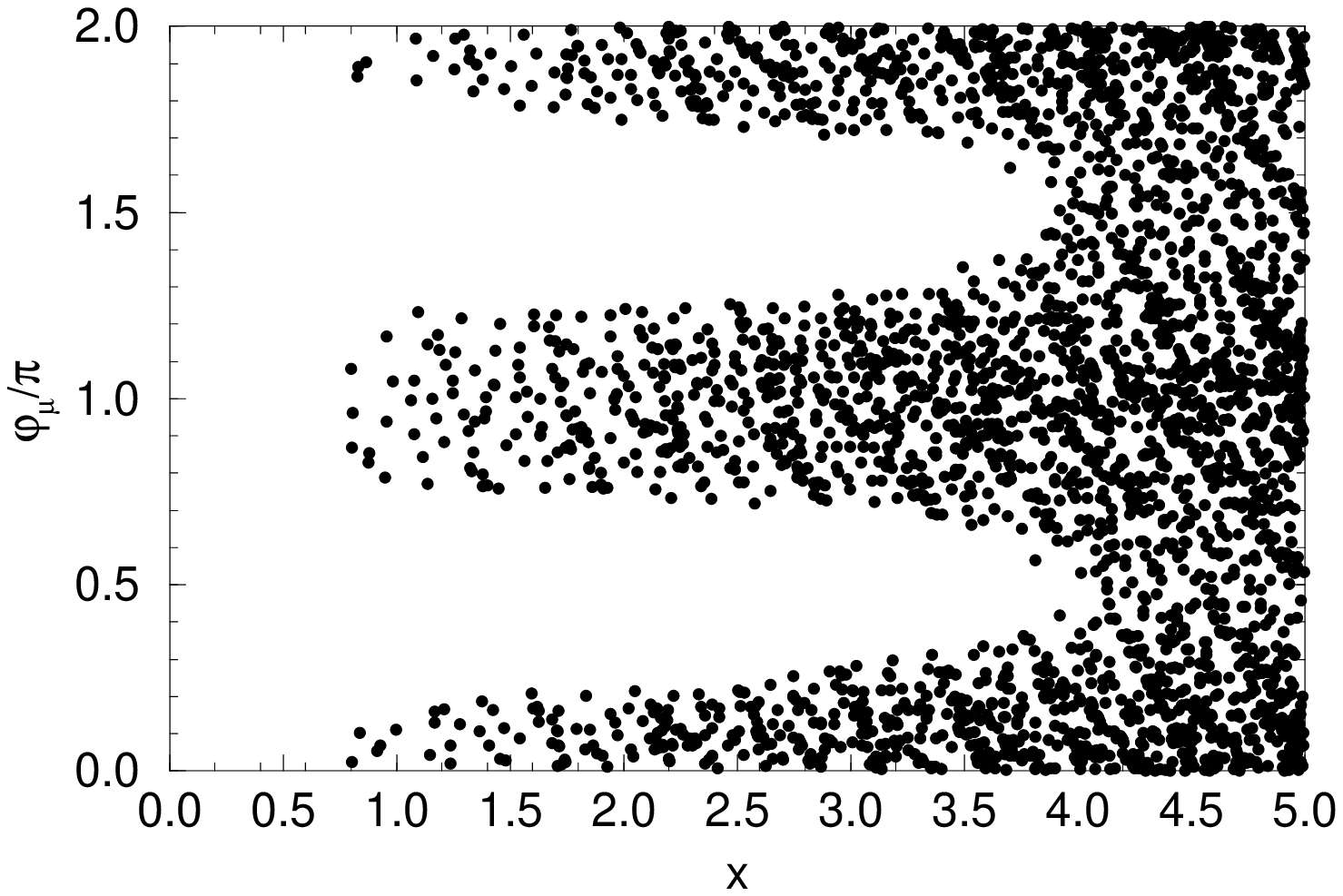}}
\vskip -5mm
\centerline{Fig.~6}
\vfill\eject

\centerline{\epsfbox{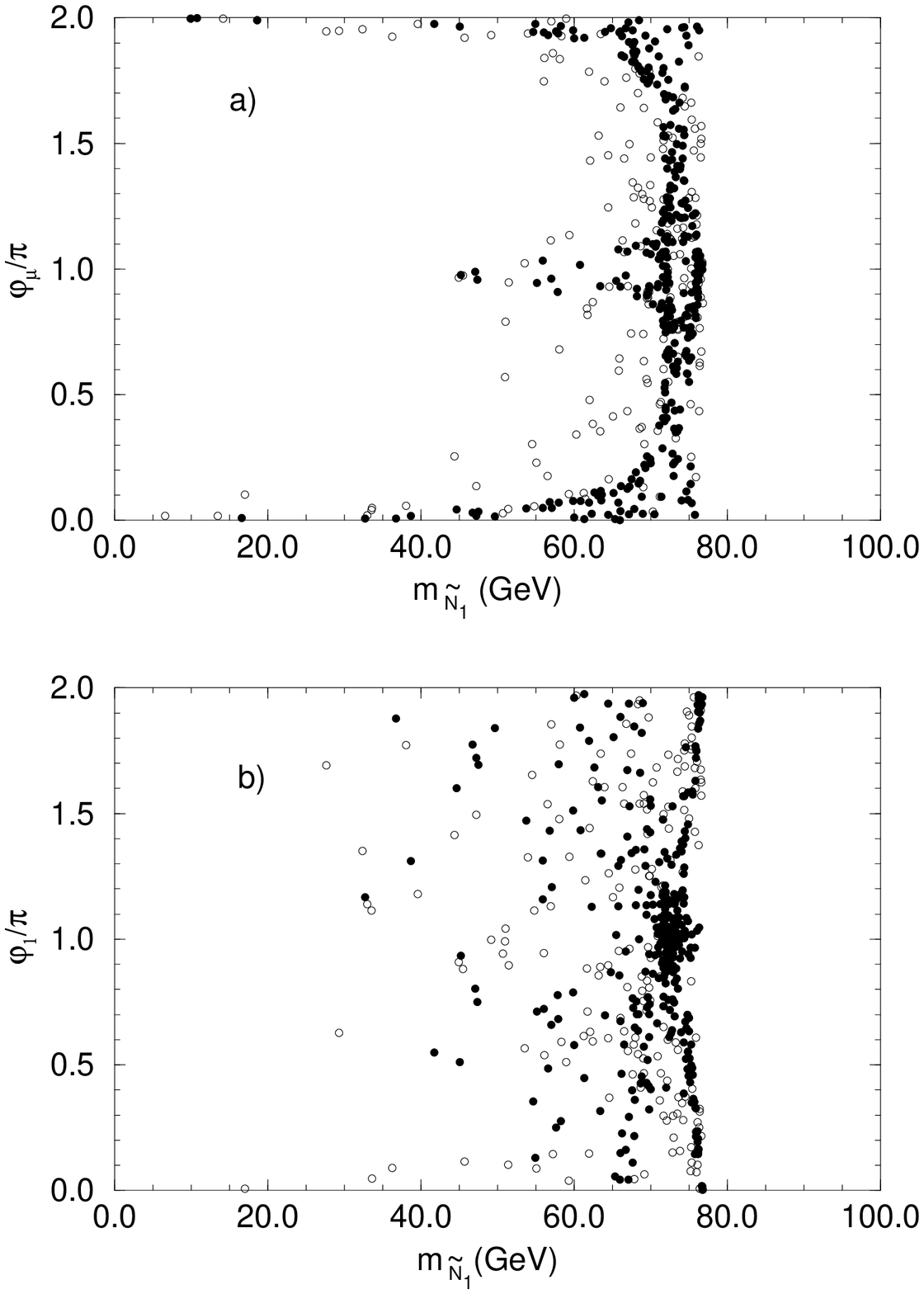}}
\vskip -5mm
\centerline{Fig.~7}
\vfill\eject

\end{document}